\documentclass[journal,transmag]{IEEEtran}
\usepackage[T1]{fontenc}  
\usepackage[english]{babel}
\usepackage{amsmath} % mathematics
\usepackage{bm} % bold mathematic expressions
\usepackage{eucal} % more beautiful calligraphic math letters
\usepackage{dsfont} % needed for more beautiful |E
\usepackage{color}
\usepackage[top=0.7in, left=0.65in]{geometry}
\usepackage{graphicx}
\usepackage{subcaption} % needed for renaming figure and table labels
\usepackage{tikz}
\usepackage{pgfplots}
\usepackage{hyperref}

\newcommand{\dd}{\mathrm{d}}
\newcommand{\Mmech}{M_\mathrm{m}}
\newcommand{\wmech}{\omega_\mathrm{m}}

\newcommand{\tend}{t_\mathrm{end}}

\newcommand{\Crr}{C_\mathrm{rr}}
\newcommand{\Cd}{C_\mathrm{d}}
\newcommand{\Jsrc}{\vec{J}_\mathrm{src}}
\newcommand{\Hpm}{\vec{H}_\mathrm{pm}}
\newcommand{\domain}{\Omega_\mathrm{d}}
\newcommand{\Crrdry}{C_\mathrm{rr,dry}}
\newcommand{\Cddry}{C_\mathrm{d,dry}}
\newcommand{\jsrc}{\bm{j}_\mathrm{src}}
\newcommand{\jpm}{\bm{j}_\mathrm{pm}}
\newcommand{\Ld}{L_\mathrm{d}}
\newcommand{\Lq}{L_\mathrm{q}}
\newcommand{\Rst}{R_\mathrm{st}}
\newcommand{\std}{\mathrm{std}}
\newcommand{\Mmax}{M_\mathrm{max}}
\newcommand{\Mmaxd}{M_\mathrm{max,d}}

\newcommand{\qoi}{q}
\newcommand{\Nc}{N_\mathrm{c}}
\newcommand{\NFE}{N_\mathrm{FE}}
\newcommand{\betaopt}{\beta_\mathrm{opt}}
\newcommand{\boundary}[1]{\Gamma_\mathrm{#1}}
\newcommand{\deltarr}{\delta_\mathrm{rr}}
\newcommand{\deltad}{\delta_\mathrm{d}}
\newcommand{\bmdeltav}{\bm{\delta}_{\bm{v}}}
\newcommand{\bmdeltap}{\bm{\delta}_{\bm{p}}}
\newcommand{\epsopt}{\epsilon_\mathrm{opt}}
\newcommand{\ppair}{N_\mathrm{pp}}

\begin{document}
	\title{Robust Optimization of a Permanent Magnet Synchronous Machine Considering Uncertain Driving Cycles}
	\author{\IEEEauthorblockN{L.~A.~M.~D'Angelo\IEEEauthorrefmark{1,2},
			Z. Bontinck\IEEEauthorrefmark{1,2}, 
			S. Sch\"ops\IEEEauthorrefmark{1,2}, and
			H. De Gersem\IEEEauthorrefmark{1,2}}
		\IEEEauthorblockA{\IEEEauthorrefmark{1}Institut f\"ur Teilchenbeschleunigung und Elektromagnetische Felder, Technische Universit\"at Darmstadt, Darmstadt, Germany}
		\IEEEauthorblockA{\IEEEauthorrefmark{2}Graduate School of Computational Engineering, Technische Universit\"at Darmstadt, Darmstadt, Germany}
	}

\markboth{CMP-836}{CMP-836}
	
\IEEEtitleabstractindextext{%
	\begin{abstract}
		This work focuses on the robust optimization of a permanent magnet (PM) synchronous machine while considering a driving cycle. The robustification is obtained by considering uncertainties of different origins. Firstly, there are geometrical uncertainties caused by manufacturing inaccuracies. Secondly, there are uncertainties linked to different driving styles. The final set of uncertainties is linked to ambient parameters such as traffic and weather conditions. The optimization goal is to minimize the PM's volume while maintaining a desired machine performance measured by the energy efficiency over the driving cycle and the machine's maximal torque. The magnetic behavior of the machine is described by a partial differential equation (PDE) and is simulated by the finite-element method employing an affine decomposition to avoid reassembling of the system of equations due to the changing PM geometry. The Sequential Quadratic Programming algorithm is used for the optimization. Stochastic collocation is applied to compute moments of stochastic quantities. The robustness of the optimized configurations is validated by a Monte Carlo sampling. It is found that the uncertainties in driving style and road conditions have significant influence on the optimal PM configuration.
	\end{abstract}
	
\begin{IEEEkeywords}
	Optimization, Permanent magnet machines, Robustness, Uncertainty
\end{IEEEkeywords}
}

\maketitle

\section{Introduction}
\IEEEPARstart{P}{ermanent} magnet (PM) synchronous machines (PMSMs) with buried magnets are widespread in electromobility due to their high efficiency and high power density. This allows a small machine for a fixed output power~\cite{Zeraoulia_2006aa},~\cite{Finken_2010aa}. PMs consist of rare-earth elements presenting a limited and expensive resource, whose extraction and recycling is environmentally polluting~\cite{Binnemans_2013aa}. Therefore, a reduction of the PM volume is desired while maintaining the performance of the machine. In practice, the machine's performance is linked to the driving conditions. Hence, the efficiency over the full driving cycle has to be taken into account, which represents an energy efficiency due to the additional time component. This quantity can be calculated by a weighting based on e.g.~energy consumption~\cite{Lazari_2012aa} or operation point statistics~\cite{Kreim 2013aa}.

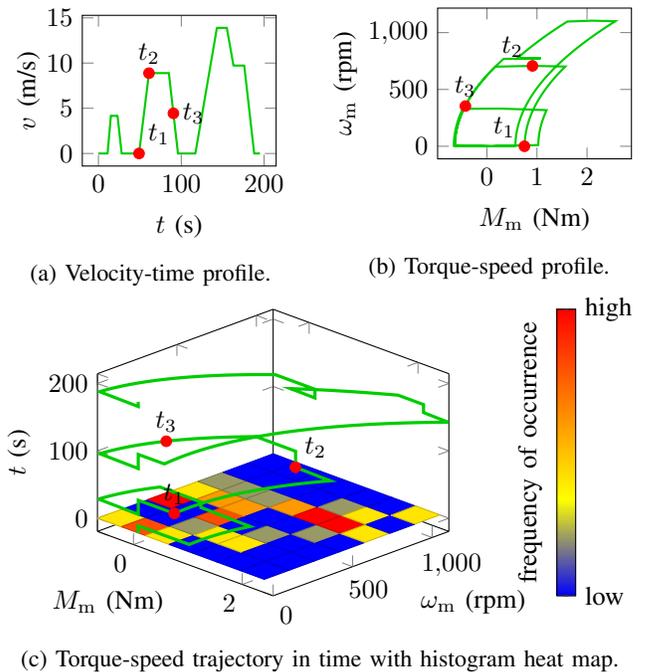
\begin{figure}[tb]
	\centering 
	\begin{subfigure}{.49\columnwidth}
		\centering 
		\begin{tikzpicture}
		\begin{axis}[width=.95\columnwidth,xlabel={$t$ (s)},ylabel={$v$ (m/s)},legend pos=north west, x label style = {at={(axis description cs:0.5,0)}}, y label style = {at={(axis description cs:0.2,0.5)}}]
		\addplot[color=green!80!black, thick] table [x={t}, y={v}, col sep=comma] {dc_vt.csv};
		\addplot[color=red, only marks] coordinates {(49,0) (61,8.88) (90.5,4.44)};
		\node at (axis cs: 49,0) [anchor=south west] {$t_1$};
		\node at (axis cs: 61,8.88) [anchor=south] {$t_2$};
		\node at (axis cs: 90.5,4.44) [anchor=west] {$t_3$};
		%\legend{$v(t)$}
		\end{axis}
		\end{tikzpicture} 
		\caption{Velocity-time profile.}
		\label{fig:dc_vt}
	\end{subfigure}
	\begin{subfigure}{.49\columnwidth}
		\begin{tikzpicture}
		\begin{axis}[width=.95\columnwidth,xlabel={$\Mmech$ (Nm)},ylabel={$\wmech$ (rpm)},legend pos=north west, x label style = {at={(axis description cs:0.5,0)}}, y label style = {at={(axis description cs:0,0.5)}}]
		\addplot[color=green!80!black, thick] table [x={M}, y={w}, col sep=comma] {dc_wM.csv};
		\addplot[color=red, only marks] coordinates {(0.75,0) (0.91,707.3) (-0.43,353.7)};
		\node at (axis cs: 0.75,0) [anchor=south east] {$t_1$};
		\node at (axis cs: 0.91,707.3) [anchor=south east] {$t_2$};
		\node at (axis cs: -0.43,353.7) [anchor=south] {$t_3$};
		%\legend{$\wmech(\Mmech)$}
		\end{axis}
		\end{tikzpicture}
		\caption{Torque-speed profile.}
		\label{fig:dc_wM}
	\end{subfigure}
	\begin{subfigure}{1\columnwidth}
		\centering 
		\begin{tikzpicture}
		\begin{axis}[width=.7\columnwidth, view={45}{30}, xlabel={$\Mmech$ (Nm)}, ylabel={$\wmech$ (rpm)}, zlabel={$t$ (s)}, x label style ={at={(axis description cs:0.1,0.1)}},  y label style = {at={(axis description cs:1.0,0.1)}}, colorbar, colorbar style={ytick = {0,0.5}, yticklabels={low,high}, width=0.25cm, ylabel={frequency of occurrence}, y label style = {at={(axis description cs:3.5,0.5)}} }]
		\addplot3[patch, shader=faceted, patch type = rectangle] coordinates {
			(-0.6604, 0 , 0.2) (-0.2226, 0, 0.2) (-0.2226, 157.9, 0.2) (-0.6604, 157.8, 0.2)
		};
		\addplot3[patch, shader=faceted, patch type = rectangle] coordinates {
			(-0.2226, 0, 0.4) (0.2151, 0, 0.4) (0.2151, 157.9, 0.4) (-0.2226, 157.9, 0.4)
		};
		\addplot3[patch, shader=faceted, patch type = rectangle] coordinates {
			(0.2151, 0, 0.2) (0.6529, 0, 0.2) (0.6529, 157.9, 0.2) (0.2151, 157.9, 0.2)
		};
		\addplot3[patch, shader=faceted, patch type = rectangle] coordinates {
			(0.6529, 0, 0) (1.091, 0, 0) (1.091, 157.9, 0) (0.6529, 157.9, 0)	
		};
		\addplot3[patch, shader=faceted, patch type = rectangle] coordinates {
			(1.091, 0, 0) (1.528, 0, 0) (1.528, 157.9, 0) (1.091, 157.9, 0)	
		};
		\addplot3[patch, shader=faceted, patch type = rectangle] coordinates {
			(1.528, 0, 0) (1.966, 0, 0) (1.966, 157.9, 0) (1.528, 157.9, 0)	
		};
		\addplot3[patch, shader=faceted, patch type = rectangle] coordinates {
			(1.966, 0, 0) (2.404, 0, 0) (2.404, 157.9, 0) (1.966, 157.9, 0)
		};
		\addplot3[patch, shader=faceted, patch type = rectangle] coordinates {
			(-0.6604, 157.9, 0) (-0.226, 157.9, 0) (-0.226, 315.8, 0) (-0.6604, 315.8, 0)	
		};
		\addplot3[patch, shader=faceted, patch type = rectangle] coordinates {
			(-0.226, 157.9, 0) (0.2151, 157.9, 0) (0.2151, 315.8, 0) (-0.226, 315.8, 0)
		};
		\addplot3[patch, shader=faceted, patch type = rectangle] coordinates {
			(0.2151, 157.9, 0.1) (0.6529, 157.9, 0.1) (0.6529, 315.8, 0.1) (0.2151, 315.8, 0.1)	
		};
		\addplot3[patch, shader=faceted, patch type = rectangle] coordinates {
			(0.6529, 157.9, 0.2) (1.091, 157.9, 0.2) (1.091, 315.8, 0.2) (0.6529, 315.8, 0.2)	
		};
		\addplot3[patch, shader=faceted, patch type = rectangle] coordinates {
			(1.091, 157.9, 0.2) (1.528, 157.9, 0.2) (1.528, 315.8, 0.2) (1.091, 315.8, 0.2)	
		};
		\addplot3[patch, shader=faceted, patch type = rectangle] coordinates {
			(1.528, 157.9, 0) (1.966, 157.9, 0) (1.966, 315.8, 0) (1.528, 315.8, 0)	
		};
		\addplot3[patch, shader=faceted, patch type = rectangle] coordinates {
			(1.966, 157.9, 0) (2.404, 157.9, 0) (2.404, 315.8, 0) (1.966, 315.8, 0)	
		};
		\addplot3[patch, shader=faceted, patch type = rectangle] coordinates {
			(-0.6604, 315.8, 0.5) (-0.2226, 315.8, 0.5) (-0.2226, 473.7, 0.5) (-0.6604, 473.7, 0.5)	
		};
		\addplot3[patch, shader=faceted, patch type = rectangle] coordinates {
			(-0.2226, 315.8, 0) (0.2151, 315.8, 0) (0.2151, 473.7, 0) (-0.2226, 473.7, 0)	
		};
		\addplot3[patch, shader=faceted, patch type = rectangle] coordinates {
			(0.2151, 315.8, 0.4) (0.6529, 315.8, 0.4) (0.6529, 473.7, 0.4) (0.2151, 473.7, 0.4)	
		};
		\addplot3[patch, shader=faceted, patch type = rectangle] coordinates {
			(0.6529, 315.8, 0) (1.091, 315.8, 0) (1.091, 473.7, 0) (0.6529, 473.7, 0)	
		};	
		\addplot3[patch, shader=faceted, patch type = rectangle] coordinates {
			(1.091, 315.8, 0.1) (1.528, 315.8, 0.1) (1.528, 473.7, 0.1) (1.091, 473.7, 0.1)	
		};
		\addplot3[patch, shader=faceted, patch type = rectangle] coordinates {
			(1.528, 315.8, 0) (1.966, 315.8, 0) (1.966, 473.7, 0) (1.528, 473.7, 0)	
		};
		\addplot3[patch, shader=faceted, patch type = rectangle] coordinates {
			(1.966, 315.8, 0) (2.404, 315.8, 0) (2.404, 473.7, 0) (1.966, 473.7, 0)	
		};
		\addplot3[patch, shader=faceted, patch type = rectangle] coordinates {
			(-0.6604, 473.7, 0.2) (-0.226, 473.7, 0.2) (-0.226, 631.6, 0.2) (-0.6604, 631.6, 0.2)	
		};
		\addplot3[patch, shader=faceted, patch type = rectangle] coordinates {
			(-0.226, 473.7, 0.3) (0.2151, 473.7, 0.3) (0.2151, 631.6, 0.3) (-0.226, 631.6, 0.3)	
		};
		\addplot3[patch, shader=faceted, patch type = rectangle] coordinates {
			(0.2151, 473.7, 0.3) (0.6529, 473.7, 0.3) (0.6529, 631.6, 0.3) (0.2151, 631.6, 0.3)	
		};
		\addplot3[patch, shader=faceted, patch type = rectangle] coordinates {
			(0.6529, 473.7, 0) (1.091, 473.7, 0) (1.091, 631.6, 0) (0.6529, 631.6, 0)	
		};
		\addplot3[patch, shader=faceted, patch type = rectangle] coordinates {
			(1.091, 473.7, 0) (1.528, 473.7, 0) (1.528, 631.6, 0) (1.091, 631.6, 0)	
		};
		\addplot3[patch, shader=faceted, patch type = rectangle] coordinates {
			(1.528, 473.7, 0.1) (1.966, 473.7, 0.1) (1.966, 631.6, 0.1) (1.528, 631.6, 0.1)	
		};
		\addplot3[patch, shader=faceted, patch type = rectangle] coordinates {
			(1.966, 473.7, 0) (2.404, 473.7, 0) (2.404, 631.6, 0) (1.966, 631.6, 0)
		};
		\addplot3[patch, shader=faceted, patch type = rectangle] coordinates {
			(-0.6604, 631.6, 0.1) (-0.2226, 631.6, 0.1) (-0.2226, 789.5, 0.1) (-0.6604, 789.5, 0.1)	
		};
		\addplot3[patch, shader=faceted, patch type = rectangle] coordinates {
			(-0.2226, 631.6, 0.1) (0.2151, 631.6, 0.1) (0.2151, 789.5, 0.1) (-0.2226, 789.5, 0.1)	
		};
		\addplot3[patch, shader=faceted, patch type = rectangle] coordinates {
			(0.2151, 631.6, 0.1) (0.6529, 631.6, 0.1) (0.6529, 789.5, 0.1) (0.2151, 789.5, 0.1)	
		};
		\addplot3[patch, shader=faceted, patch type = rectangle] coordinates {
			(0.6529, 631.6, 0.3) (1.091, 631.6, 0.3) (1.091, 789.5, 0.3) (0.6529, 789.5, 0.3)	
		};
		\addplot3[patch, shader=faceted, patch type = rectangle] coordinates {
			(1.091, 631.6, 0.5) (1.528, 631.6, 0.5) (1.528, 789.5, 0.5) (1.091, 789.5, 0.5)	
		};
		\addplot3[patch, shader=faceted, patch type = rectangle] coordinates {
			(1.528, 631.6, 0.5) (1.966, 631.6, 0.5) (1.966, 789.5, 0.5) (1.528, 789.5, 0.5)	
		};
		\addplot3[patch, shader=faceted, patch type = rectangle] coordinates {
			(1.966, 631.6, 0.2) (2.404, 631.6, 0.2) (2.404, 789.5, 0.2) (1.966, 789.5, 0.2)	
		};
		\addplot3[patch, shader=faceted, patch type = rectangle] coordinates {
			(-0.6604, 789.5, 0) (-0.226, 789.5, 0) (-0.226, 947.4, 0) (-0.6604, 947.4, 0)	
		};
		\addplot3[patch, shader=faceted, patch type = rectangle] coordinates {
			(-0.226, 789.5, 0) (0.2151, 789.5, 0) (0.2151, 947.4, 0) (-0.226, 947.4, 0)	
		};
		\addplot3[patch, shader=faceted, patch type = rectangle] coordinates {
			(0.2151, 789.5, 0) (0.6529, 789.5, 0) (0.6529, 947.4, 0) (0.2151, 947.4, 0)	
		};
		\addplot3[patch, shader=faceted, patch type = rectangle] coordinates {
			(0.6529, 789.5, 0) (1.091, 789.5, 0) (1.091, 947.4, 0) (0.6529, 947.4, 0)	
		};
		\addplot3[patch, shader=faceted, patch type = rectangle] coordinates {
			(1.091, 789.5, 0.1) (1.528, 789.5, 0.1) (1.528, 947.4, 0.1) (1.091, 947.4, 0.1)	
		};
		\addplot3[patch, shader=faceted, patch type = rectangle] coordinates {
			(1.528, 789.5, 0.2) (1.966, 789.5, 0.2) (1.966, 947.4, 0.2) (1.528, 947.4, 0.2)	
		};
		\addplot3[patch, shader=faceted, patch type = rectangle] coordinates {
			(1.966, 789.5, 0) (2.404, 789.5, 0)	(2.404, 947.4, 0) (1.966, 947.4, 0)
		};
		\addplot3[patch, shader=faceted, patch type = rectangle] coordinates {
			(-0.6604, 947.4, 0) (-0.2226, 947.4, 0) (-0.2226, 1105, 0) (-0.6604, 1105, 0)	
		};
		\addplot3[patch, shader=faceted, patch type = rectangle] coordinates {
			(-0.2226, 947.4, 0) (0.2151, 947.4, 0) (0.2151, 1105, 0) (-0.2226, 1105, 0)	
		};
		\addplot3[patch, shader=faceted, patch type = rectangle] coordinates {
			(0.2151, 947.4, 0) (0.6529, 947.4, 0) (0.6529, 1105, 0) (0.2151, 1105, 0)	
		};
		\addplot3[patch, shader=faceted, patch type = rectangle] coordinates {
			(0.6529, 947.4, 0) (1.091, 947.4, 0) (1.091, 1105, 0) (0.6529, 1105, 0)	
		};
		\addplot3[patch, shader=faceted, patch type = rectangle] coordinates {
			(1.091, 947.4, 0) (1.528, 947.4, 0) (1.528, 1105, 0) (1.091, 1105, 0)	
		};
		\addplot3[patch, shader=faceted, patch type = rectangle] coordinates {
			(1.528, 947.4, 0) (1.966, 947.4, 0) (1.966, 1105, 0) (1.528, 1105, 0)	
		};
		\addplot3[patch, shader=faceted, patch type = rectangle] coordinates {
			(1.966, 947.4, 0.2) (2.404, 947.4, 0.2) (2.404, 1105, 0.2) (1.955, 1105, 0.2)	
		};
		\addplot3[color=green!80!black, very thick] table [x={M}, y={w}, z={t}, col sep=comma] {dc_3d.csv};
		\addplot3[color=red, only marks] coordinates {(0.75,0,49) (0.91,707.3,61) (-0.43,353.7,90.5)};
		\node at (axis cs: 0.75,0,49) [anchor=south] {$t_1$};
		\node at (axis cs: 0.91,707.3,61) [anchor=south west] {$t_2$};
		\node at (axis cs: -0.43,353.7,90.5) [anchor=south] {$t_3$};
		\end{axis}
		\end{tikzpicture}
		\caption{Torque-speed trajectory in time with histogram heat map.}
		\label{fig:dc_3d}
	\end{subfigure}
	\caption{Different representations of the UDC. For three points in time $t_1,t_2,t_3$, the corresponding positions in the representations have been marked.}
	\label{fig:urbanDC}
\end{figure}

The machine performance suffers from uncertainties occurring in the PM characteristics due to manufacturing inaccuracies~\cite{Coenen_2012aa}, and in the driving cycle because of changing traffic, weather situations and driving style, e.g.~stronger accelerations. Thus, the optimal PM configuration has to be robust, i.e.~even under these slight variations the machine should still perform at least as specified. 

To evaluate the performance of the machine, the finite-element method (FEM) is utilized. Ultimately, this results in a nonlinear robust optimization problem constrained by a partial differential equation (PDE). This overall problem has extremely high computational demands, because the PDE has to be solved once anew for every optimization step and, even more problematic, every stochastic quantity adds a dimension to the problem.
Hence, the choice of the optimization method and the procedure to calculate the stochastic quantities are crucial. For this reason, metaheuristic optimization methods -- as they are employed e.g.~in~\cite{Degano_2014aa} -- are unsuitable for this problem, as they converge slowly and thus need many optimization steps. %~\cite{Yang_2010aa}. 
In contrast, gradient-based methods typically converge fast, provided that the gradients of the objective and constraint function are given with a sufficient accuracy. Although this can be cumbersome for complicated functions, such preliminary computations are rewarded by a low iteration count of the embracing optimization procedure~\cite{Lass_2017aa}. The comparably large number of uncertain parameters is efficiently treated by stochastic collocation on a Clenshaw-Curtis sparse grid~\cite{Bungartz_2004aa,Nobile_2008aa}. The performance is further improved using an affine decomposition, i.e.~the geometry is divided into a fixed region and a region with a changing geometry which is deformed by an affine map~\cite{Rozza_2008aa}.

This work is structured as follows. First, the concept of driving cycles and how to compute the performance of a machine with respect to a driving cycle as well as the modeling of uncertainties is explained in Sec.~\ref{sec:drivingcycle}. Then, the PMSM model and its computation are introduced in Sec.~\ref{sec:pmsm}. Sec.~\ref{sec:opt} presents the robust optimization problem and method. It also discusses how to compute the stochastic quantities. Finally, the results and conclusions of this work are discussed in Sec.~\ref{sec:res}.

\section{Driving Cycle}\label{sec:drivingcycle}
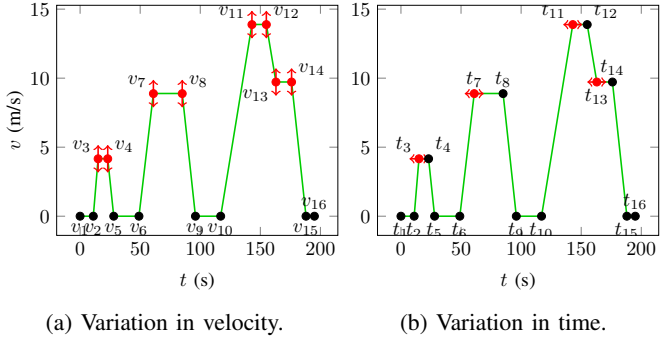
\begin{figure}[tb]
	\centering 
	\begin{subfigure}{.49\columnwidth}
		\centering 
		\begin{tikzpicture}[scale=.75]
		\begin{axis}[width=1.5\columnwidth, xlabel={$t$ (s)},ylabel={$v$ (m/s)},legend pos=north west, x label style = {at={(axis description cs:0.5,0)}}, y label style = {at={(axis description cs:0.1,0.5)}}]
		\addplot[color=green!80!black, thick] table [x={t}, y={v}, col sep=comma] {dc_vt_coarse.csv};
		\addplot[color=black,only marks] coordinates {(0,0) (11,0) (28,0) (49,0) (96,0) (117,0) (188,0) (195,0)};
		\addplot[color=red,only marks] coordinates {(15,4.16) (23,4.16) (61,8.88) (85,8.88) (143,13.88) (155,13.88) (163,9.72) (176,9.72)};
		\draw[<->,thick,red] (axis cs: 15,3.16) -- (axis cs: 15,5.16);
		\draw[<->,thick,red] (axis cs: 23,3.16) -- (axis cs: 23,5.16);
		\draw[<->,thick,red] (axis cs: 61,7.88) -- (axis cs: 61,9.88);
		\draw[<->,thick,red] (axis cs: 85,7.88) -- (axis cs: 85,9.88);
		\draw[<->,thick,red] (axis cs: 143,12.88) -- (axis cs: 143,14.88);
		\draw[<->,thick,red] (axis cs: 155,12.88) -- (axis cs: 155,14.88);
		\draw[<->,thick,red] (axis cs: 163,8.72) -- (axis cs: 163,10.72);
		\draw[<->,thick,red] (axis cs: 176,8.72) -- (axis cs: 176,10.72);
		\node at (axis cs: 0,0) [anchor=north] {$v_1$};
		\node at (axis cs: 11,0) [anchor=north] {$v_2$};
		\node at (axis cs: 15,4.16) [anchor=south east] {$v_3$};
		\node at (axis cs: 23,4.16) [anchor=south west] {$v_4$};
		\node at (axis cs: 28,0) [anchor=north] {$v_5$};
		\node at (axis cs: 49,0) [anchor=north] {$v_6$};
		\node at (axis cs: 61,8.88) [anchor=south east] {$v_7$};
		\node at (axis cs: 85,8.88) [anchor=south west] {$v_8$};
		\node at (axis cs: 96,0) [anchor=north] {$v_9$};
		\node at (axis cs: 117,0) [anchor=north] {$v_{10}$};
		\node at (axis cs: 143,13.88) [anchor=south east] {$v_{11}$};
		\node at (axis cs: 155,13.88) [anchor=south west] {$v_{12}$};
		\node at (axis cs: 163,9.72) [anchor=north east] {$v_{13}$};
		\node at (axis cs: 176,9.72) [anchor=south west] {$v_{14}$};
		\node at (axis cs: 188,0) [anchor=north] {$v_{15}$};
		\node at (axis cs: 195,0) [anchor=south] {$v_{16}$};
		\end{axis}
		\end{tikzpicture}
		\caption{Variation in velocity.}
		\label{fig:dc_with_controlpoints_A}
	\end{subfigure}
	\begin{subfigure}{.49\columnwidth}
		\centering 
		\begin{tikzpicture}[scale=.75]
		\begin{axis}[width=1.5\columnwidth,xlabel={$t$ (s)},ylabel={},legend pos=north west, x label style = {at={(axis description cs:0.5,0)}}, y label style = {at={(axis description cs:0.1,0.5)}}]
		\addplot[color=green!80!black, thick] table [x={t}, y={v}, col sep=comma] {dc_vt_coarse.csv};
		\addplot[color=black,only marks] coordinates {(0,0) (11,0) (23,4.16) (28,0) (49,0) (85,8.88) (96,0) (117,0) (155,13.88) (176,9.72) (188,0) (195,0)};
		\addplot[color=red,only marks] coordinates {(15,4.16)  (61,8.88) (143,13.88)  (163,9.72)};
		\draw[<->,thick,red] (axis cs: 7,4.16) -- (axis cs: 23,4.16);
		\draw[<->,thick,red] (axis cs: 53,8.88) -- (axis cs: 69,8.88);
		\draw[<->,thick,red] (axis cs: 135,13.88) -- (axis cs: 151,13.88);
		\draw[<->,thick,red] (axis cs: 155,9.72) -- (axis cs: 171,9.72);
		\node at (axis cs: 0,0) [anchor=north] {$t_1$};
		\node at (axis cs: 11,0) [anchor=north] {$t_2$};
		\node at (axis cs: 15,4.16) [anchor=south east] {$t_3$};
		\node at (axis cs: 23,4.16) [anchor=south west] {$t_4$};
		\node at (axis cs: 28,0) [anchor=north] {$t_5$};
		\node at (axis cs: 49,0) [anchor=north] {$t_6$};
		\node at (axis cs: 61,8.88) [anchor=south] {$t_7$};
		\node at (axis cs: 85,8.88) [anchor=south] {$t_8$};
		\node at (axis cs: 96,0) [anchor=north] {$t_9$};
		\node at (axis cs: 117,0) [anchor=north] {$t_{10}$};
		\node at (axis cs: 143,13.88) [anchor=south east] {$t_{11}$};
		\node at (axis cs: 155,13.88) [anchor=south west] {$t_{12}$};
		\node at (axis cs: 163,9.72) [anchor=north] {$t_{13}$};
		\node at (axis cs: 176,9.72) [anchor=south] {$t_{14}$};
		\node at (axis cs: 188,0) [anchor=north] {$t_{15}$};
		\node at (axis cs: 195,0) [anchor=south] {$t_{16}$};
		\end{axis}
		\end{tikzpicture}
		\caption{Variation in time.}
		\label{fig:dc_with_controlpoints_B}
	\end{subfigure}
	\caption{UDC as piece-wise linear spline with sixteen control points. Uncertainties can be implemented by deviating a subset of control points vertically (a) or horizontally (b) or both.}
	\label{fig:dc_with_controlpoints}
\end{figure}
%\subsection{Urban Driving Cycle}
The Urban Driving Cycle (UDC) is part of the New European Driving Cycle specified in~\cite{UN_2005aa}, where it is defined by its velocity-time profile~$v(t)$ depicted in Fig.~\ref{fig:dc_vt}. From this, the torque-speed profile~$(\Mmech,\wmech)$ (Fig.~\ref{fig:dc_wM}) is obtained by mechanical laws. However, this representation lacks the information of how long a specific operation point~$(\Mmech,\wmech)$ is operated, which is needed to compute the energy balance over the UDC. Former works partitioned the torque-speed plane and performed some sort of weighting~\cite{Lazari_2012aa}-\cite{Kreim 2013aa}. For instance, the torque-speed plane can be tiled into a $8\times8$ grid. Then, it is counted how many points of the torque-speed curve fall in each square while sampling the curve equidistantly. As a result, a histogram is obtained which is visualized as a heat map on the grid (see Fig.~\ref{fig:dc_3d}).
In contrast, this work considers the time information by including the time axis as third component, leading to the UDC trajectory in Fig.~\ref{fig:dc_3d}. In this way, the energy and thus the efficiency over the UDC can be calculated by integrating over this torque-speed trajectory in time using efficient Gaussian quadrature. The efficiency is
\begin{equation}
	\mathcal{E}(\bm{p}) = \frac{ \int\limits_{t_0}^{\tend} \wmech(t) \Mmech(t) \, \dd t }{\int\limits_{t_0}^{\tend} \wmech(t) \Mmech(t) + m \Rst I^2(t,\bm{p}) \, \dd t },
	\label{eq:energy_efficiency}
\end{equation}
where $\bm{p}$ collects parameters of the PM geometry, $m$ is the number of phases, $\Rst$ is the stator resistance in $\Omega$ and $I$ is the stator current in A given by the nonlinear relation 
\begin{equation*}
	\frac{\Mmech(t,\bm{p})}{\ppair m} =  I \sin(\beta) \left( \Phi_0(\bm{p}) + \left( \Ld(\bm{p})-\Lq(\bm{p}) \right) I \cos(\beta) \right) 
\end{equation*}
with $\ppair$ as pole pair number, $\Phi_0$ as magnetic flux at no-load in Wb, $\Ld$ as direct-axis inductance in H, $\Lq$ as quadrature-axis inductance in H and $\beta = \betaopt(t,\bm{p})$ as the optimal current phase angle leading to the highest possible torque output~\cite{Pahner_1998aa}.

To model uncertainties in the driving cycle, the curve~$v(t)$ is interpreted as piece-wise linear splines with sixteen control points. Three scenarios arise: 
In scenario A, the control points~$\bm{v}$ are deviated vertically in velocity (Fig.~\ref{fig:dc_with_controlpoints_A}) while assuming them to be uniformly distributed,
\begin{equation}
	\bm{v}(\theta) \in \mathcal{U}\left( (1-\bm{\delta}_{\bm{v}}) \bar{\bm{v}}, (1+\bm{\delta}_{\bm{v}}) \bar{\bm{v}} \right)
	\label{eq:vertical_shift}
\end{equation}
with $\mathcal{U}(a,b)$ standing for the uniform distribution in the interval $[a,b]$, $\theta$ as quantity representing the stochastic nature, $\bm{\delta}_{\bm{v}}$ as maximal relative deviation and $\bar{\bm{v}}$ as the control point's mean values. This corresponds e.g.~to drivers ignoring speed limits.

In scenario B, the control points are shifted horizontally in time (Fig.~\ref{fig:dc_with_controlpoints_B}) by choosing the variations as 
\begin{equation}
	\delta t_i(\theta) = \left\lbrace \begin{array}{ll}
	\alpha \tau(\theta) (t_i - t_{i-1}) & \text{for $\tau(\theta) < 0$,} \\
	\alpha \tau(\theta) (t_{i+1} - t_i) & \text{otherwise}
	\end{array} \right.
	\label{eq:horizontal_shift}
\end{equation}
with $\tau(\theta) \in \mathcal{U}(-1,1)$ and some constant factor $\alpha \in (0,1)$ controlling the strength of the deviation.
A combination of both scenarios is also possible (scenario A+B). These three scenarios can mimic changing traffic situations or different driving habits. 

Instead of manipulating the velocity-time profile itself, one can also variate the rolling resistance coefficient~$C_{\mathrm{rr}}$ and the drag coefficient~$C_{\mathrm{d}}$, which are used to compute~$\Mmech(t)$ from~$v(t)$ (scenario C). In this way, uncertain weather and road conditions can be simulated, as~$\Crr$ can increase by a factor up to $\deltarr = 1.3$ on a wet road~\cite{Ejsmont_2015aa}, and~$\Cd$ by a factor up to $\deltad = 1.2$ due to rain drops contaminating the vehicle surface~\cite{Gaylard_2017aa}. It is assumed that these mechanical coefficients are uniformly distributed,
\begin{align}
	\Crr(\theta) &\in \mathcal{U}\left( \Crrdry, \deltarr \, \Crrdry \right), \label{eq:Crr} \\
	\Cd(\theta) &\in \mathcal{U} \left( \Cddry, \deltad \, \Cddry \right), \label{eq:Cd}
\end{align}
where $\Crrdry$ and $\Cddry$ represent the coefficients for dry weather conditions.

Please note that one can conveniently incorporate arbitrarily complicated driving cycles due to the usage of splines. If measured data are available, then one may use a Karhunen-Loève expansion to obtain a mean cycle and perturbations~\cite{Schwab_2006aa}.

\section{Computation of PMSMs}\label{sec:pmsm}
\begin{figure}[tbp]
	\centering 
	\includegraphics[width=1\columnwidth]{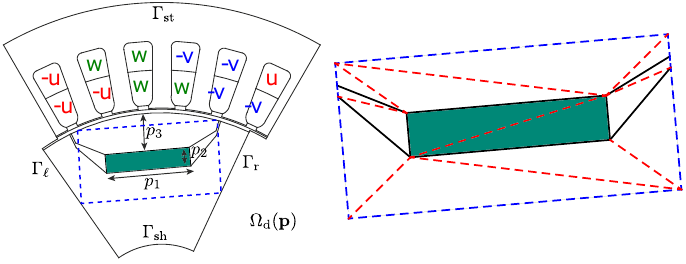}
	\caption{Left: Single pole of PMSM. The PM is marked in green. The three-phase current system is denoted with letters $\lbrace u,v,w \rbrace$, where a negative sign indicates the current flowing in the opposite direction. The geometry is subdivided following an affine decomposition. Figure adapted from~\cite{Bontinck_2018aa}.}
	\label{fig:pmsm_geometry}
\end{figure}
To save computational effort, one typically considers a two-dimensional (2D) cross-section of one single pole of the PMSM (see Fig.~\ref{fig:pmsm_geometry}). The PM (marked in green) is characterized by its width~$p_1$, its height~$p_2$ and its distance to the rotor surface~$p_3$, which are collected in a vector~$\bm{p}$. The variations due to manufacturing inaccuracies are modeled by a uniform distribution
\begin{equation}
	\bm{p}(\theta) \in \mathcal{U}\left( \bar{\bm{p}}-\bm{\delta}_{\bm{p}}, \bar{\bm{p}}+\bm{\delta}_{\bm{p}} \right),
	\label{eq:PM_uncertainty}
\end{equation}
where $\bar{\bm{p}}$ contains the mean values of the parameters in mm and $\bm{\delta}_{\bm{p}}$ is the maximal deviation in mm.

The magnetic behavior of the PMSM is described by
\begin{equation}
	\nabla \times \left( \nu( \bm{p} ) \nabla \times \vec{A}( \bm{p} ) \right) = \Jsrc - \nabla \times \Hpm( \bm{p} ) \quad \text{on $\domain(\bm{p})$}
	\label{eq:potential_formulation}
\end{equation}
with $\nu$ as reluctivity in m/H, $\vec{A}$ as magnetic vector potential (MVP) in Wb/m, $\Jsrc$ as source current density in A/m$^2$ and $\Hpm$ as the PM's source field in A/m. The MVP is discretized by $\vec{A}(\bm{p}) \approx \sum_{j=1}^{\NFE} u_j(\bm{p}) \frac{N_j}{\ell_z} \vec{e}_z$, 
where $\NFE$ is the number of degrees of freedom, $u_j$ are unknown coefficients, $N_j = N_j(x,y)$ are the 2D Cartesian nodal shape functions, $\ell_z$ is the length of the machine and $\vec{e}_z$ is the Cartesian unit vector in $z$-direction. On the stator and shaft boundaries $\boundary{st}$ and $\boundary{sh}$, homogeneous Dirichlet boundary conditions (BCs) are imposed, while antiperiodic BCs are set on the left and right side of the pole, $\boundary{\ell}$ and $\boundary{r}$ (cf.~Fig.~\ref{fig:pmsm_geometry}).
The FEM and the Ritz-Galerkin ansatz %~\cite{Brenner_2008aa}, 
lead to the system of equations 
\begin{equation}
	\bm{K}_\nu (\bm{p}) \bm{u}( \bm{p} ) = \jsrc + \jpm(\bm{p}),
	\label{eq:system}
\end{equation}
where $\bm{K}_\nu$ is the system matrix, $\jsrc$ and $\jpm$ form the right-hand side and $\bm{u}$ collects the coefficients of the discretized MVP. From $\bm{u}$, the quantities of interest (QoI) can be calculated in post-processing. To simplify notation, we state $\mathcal{E}(\bm{u}(\bm{p})) = \mathcal{E}(\bm{p})$ and $\Mmax(\bm{u}(\bm{p})) = \Mmax(\bm{p})$. 

As the system matrix as well as the right-hand side depend on the PM geometry $\bm{p}$, they would have to be assembled anew in every optimization step, which is critical as the matrix assembly requires a high computational effort. To circumvent this, an affine decomposition is employed~\cite{Rozza_2008aa}. The computational domain is subdivided in a fixed region and a region with a changing geometry (Fig.~\ref{fig:pmsm_geometry}). The system matrix and right-hand side in~\eqref{eq:system} are decomposed into fixed submatrices and subvectors multiplied by coefficients that fully inherit the dependence on $\bm{p}$. In this way, the submatrices and subvectors need to be assembled only once on beforehand, while the weighting coefficients are promptly computed by evaluating an analytical formula every time the geometry changes. 
%Furthermore, this allows to use (linear) model order reduction techniques to further speed up the simulation~\cite{Bontinck_2018ab}.
%Then, the system matrix can be decomposed into 
%\begin{equation}
%	\bm{K}_\nu(\bm{p}) = \bm{K}_\nu^0 + \sum\limits_{\ell = 1}^{N_L} \vartheta^\ell(\bm{p}) \bm{K}^\ell_\nu,
%\end{equation}
%where $\bm{K}_\nu^0$ represents the system matrix in the fixed region, $\bm{K}_\nu^\ell$ system matrices in triangular subdomains of the varying region and $\vartheta^\ell(\bm{p})$ are analytically computable coefficients that fully inherit the dependence on the PM geometry. An analogous decomposition is made for the vector $\jpm$.

The FEM solution $\bm{u}$ is used to calculate the parameters $\Ld$, $\Lq$ and $\Phi_0$ with help of the loading method introduced in~\cite{Rahman_1991aa}. Furthermore, the stator current $I$ and ultimately the energy efficiency~\eqref{eq:energy_efficiency} and torque are calculated.

\section{Robust Optimization}\label{sec:opt}
Due to the uncertainties in the driving cycle and the PM geometry, a robust minimal PM configuration is desired. To this end, the expectation value~$\mathds{E}[\cdot]$ and the standard deviation~$\std[\cdot]$ of the QoIs are needed and the standard deviation is weighted with a so-called risk aversion parameter~$\lambda$, yielding the robust optimization problem, \cite{Bontinck_2018aa} 
\begin{equation*} \left\lbrace 
	\begin{array}{lll}
		\min\limits_{\bar{\bm{p}}} & J(\bm{p}) = \bar{p}_1 \bar{p}_2 + \lambda \, \std \left[ p_1 p_2 \right] & \\
		\text{s.t.} & \mathcal{E}_d - \mathds{E}\left[ \mathcal{E} (\bm{p}) \right] + \lambda \, \std \left[ \mathcal{E} ( \bm{p}) \right] & \leq 0, \\
		& \Mmaxd - \mathds{E}\left[ \Mmax(\bm{p}) \right] + \lambda \, \std \left[ \Mmax(\bm{p}) \right] & \leq 0, \\
		& G(\bm{p}) & \leq 0.
	\end{array}\right. 
\end{equation*}
Here, $\mathcal{E}_d$ and $\Mmaxd$ are the desired energy efficiency and maximal mechanical torque, respectively, and $G(\bm{p})$ is a function considering geometrical constraints.

\begin{table}
	\centering 
	\caption{Comparison of needed evaluations of the PDE~\eqref{eq:potential_formulation} for full tensor grids and Smolyak sparse grids.}
	\begin{tabular}{c|c|c|c}
		\textbf{Scenario} & \textbf{\#Parameters} & \textbf{\#Evaluations (full)} & \textbf{\#Evaluations (sparse)} \\ \hline 
		C & 5 & 3125 & 241 \\
		B & 7 & 78125 & 589 \\
		A & 11 & 48,828,125 & 2069 \\
		A+B & 15 & $> 3 \cdot 10^{10}$ & 5021 \\ \hline 
	\end{tabular}
	\label{tab:evaluations}
\end{table}

In the process of stochastic collocation, the probabilistic integrals defining the expectation value and standard deviation of some QoI $\qoi$ are approximated by
\begin{equation}
	\mathds{E}[\qoi] \approx \sum\limits_{k=1}^{\Nc} w_k \qoi (\bm{z}_k), \quad \std[\qoi] \approx \sqrt{ \sum\limits_{k=1}^{\Nc} w_k \qoi^2 (\bm{z}_k) }, 
\end{equation}
where $\Nc$ is the number of collocation points, $w_k$ are weights and $\bm{z}_k$ are collocation points. The QoI is written as an expansion in orthogonal polynomials. The expansion coefficients are computed with a tensor product quadrature set. However, this method suffers from the curse of dimensionality, i.e.~the number of needed evaluations rises exponentially with the number of uncertain parameters~\cite{Bungartz_2004aa}, such that this method is only usable for scenario C at the very most, cf.~third column in Tab.~\ref{tab:evaluations}. Instead, Smolyak sparse grids are employed, which represent subsets of the full tensor grids~\cite{Bungartz_2004aa}. Furthermore, Clenshaw-Curtis knots are used to lower the number of needed PDE evaluations further~\cite{Nobile_2008aa}. As a result, the curse of dimensionality is significantly weakened, cf.~fourth column in Tab.~\ref{tab:evaluations}.

Eventually, the robust optimization problem is solved with Sequential Quadratic Programming, which is a standard gradient-based method to find the optimum of a constrained nonlinear optimization problem. %~\cite{Geiger_2002aa}. 
The gradients of the objective and constraint functions are determined manually and semi-analytically by symbolic computing and have been validated against numerical gradients computed with central finite differences. 
%On a side note, finite differences could be generally used to supply the gradient information, but this would demand at least two more evaluation steps of~\eqref{eq:potential_formulation} for every optimization step in case of finite differences of first order or even three more evaluations for more accuracy with central differences. 
%To ensure global convergence, the Hessian matrix is approximated using the Broyden-Fletcher-Goldfarb-Shanno update procedure.%~\cite{Geiger_2002aa}.

\section{Results \& Conclusions}\label{sec:res}
\begin{figure}[tbp]
	\centering 
	\begin{tikzpicture}[scale=.7] 
	\begin{axis}[view={0}{90}, xlabel={$I$ (A)},ylabel={$\wmech$ (rpm)},colorbar, colorbar style = {ylabel = ${\epsopt}-\epsilon_0$, y label style = {at={(axis description cs:5.5,0.5)}}}, colormap name = viridis, legend style={fill=white!80!black}] 
	\addplot3[surf,mesh/rows=40,mesh/ordering=colwise,shader=interp, forget plot] file {eps_difference_j.dat}; 
	\addplot[color=red, thick] table [x={I}, y={wmech}, col sep=comma] {udc_a.csv};
	\addplot[color=white, thick, dashed] table [x={I}, y={wmech}, col sep=comma] {udc_j.csv};
	\addlegendentry{Initial};
	\addlegendentry{Scenario C};
	\end{axis} 
	\end{tikzpicture}
	\caption{Comparison of the UDCs and efficiency between the initial and the optimized PM configuration according to scenario C: The UDCs are plotted within the $(I,\wmech)$ plane. The color depicts the absolute point-wise efficiency differences between the two cases.}
	\label{fig:efficiency}
\end{figure}
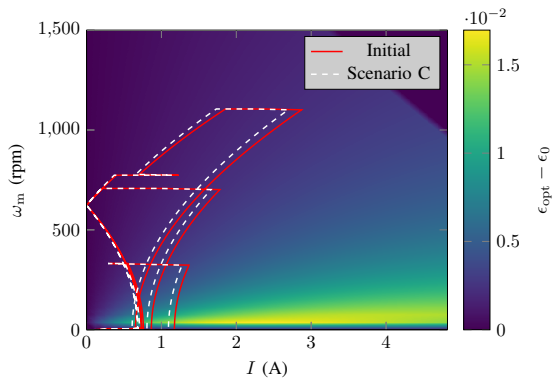
\begin{table}
	\centering 
	\caption{Resulting PM configurations and success rates for the different scenarios in Sec.~\ref{sec:drivingcycle}.}
	\begin{tabular}{l|c|c|c|c|c|c}
		\textbf{Opt. Scenario} & \textbf{Init.} & \textbf{Nom.} & \textbf{A} & \textbf{B} & \textbf{A+B} & \textbf{C} \\ \hline  
		\textbf{Size} (mm$^2$) & 133 & 63.33 & 82.70 & 86.69 & 83.46 & 100.09 \\ 
		\textbf{SR} (\%) w.r.t.~A & 42.96 & 42.88 & 99.99 & 100 & 100 & 100 \\
		\textbf{SR} (\%) w.r.t.~B & 89.82 & 83.59 & 100 & 100 & 100 & 100 \\
		\textbf{SR} (\%) w.r.t.~A+B & 74.37 & 70.89 & 100 & 100 & 100 & 100 \\
		\textbf{SR} (\%) w.r.t.~C & 0 & 0 & 20.45 & 38.77 & 23.10 & 91.14 \\ \hline 
	\end{tabular}
\label{tab:results}
\end{table}
The robust optimization of a PMSM considering an uncertain driving cycle and PM geometry is now applied to the PMSM model of Sec.~\ref{sec:pmsm}, where an initial PM configuration with a cross-sectional area $S_0 = 133\,$mm$^2$ is chosen. The desired energy efficiency is set to the value of the initial configuration, and the desired maximal torque is adapted to the maximal occurring torque in the UDC. The deviation parameters are set to $\bmdeltav = 0.2$ and $\alpha = 0.78$ for the control point's vertical~\eqref{eq:vertical_shift} and horizontal shift~\eqref{eq:horizontal_shift}, respectively, $\deltarr = 1.3$ and $\deltad = 1.2$ for the mechanical coefficients~\eqref{eq:Crr}-\eqref{eq:Cd}, and $\bmdeltap = 0.2\,$mm for the PM characteristics~\eqref{eq:PM_uncertainty}.
To validate the robustness of the found optimal PM configuration, Monte Carlo sampling is used, where the success rate (SR) is calculated as the ratio of the number of successful samples, i.e.~the samples fulfilling the desired energy efficiency and maximal torque, and the total number of samples, which is here chosen as 10,000.

The optimal PM configuration sizes found for all scenarios are listed in Tab.~\ref{tab:results}. Additionally, the initial configuration and the optimal configuration of the nominal optimization problem, i.e.~without uncertainties, are listed in the second and third column, respectively. The robustness of every configuration has been validated not only for the scenario it has been optimized for, but for every scenario. The resulting success rates are shown in Tab.~\ref{tab:results}. It is observed that the optimal PM configurations for the scenarios A, B and A+B are very similar and robust with respect to the scenario it has been optimized for, while the initial configuration as well as the nominally optimized configuration are never robust. Also, the optimal configurations for A, B and A+B are for each other robust. Thus it seems that it does not matter in which direction the control points of the UDC are deviated to model uncertain driving cycles. The largest PM configuration is returned for scenario C, for which its own robustness validation scores only a SR of $91.14\,\%$, and for which the optimized configurations for the other scenarios perform very badly. To compare the initial PM configuration with the optimized one for scenario C, Fig.~\ref{fig:efficiency} shows the resulting UDCs in the current-speed plane, since the maximal torque depends on the PM configuration and the maximal current does not. It is observed that the optimized PMSM (white curve) needs less current to achieve the same speeds as the initial one (red curve). Moreover, the operation-point-wise efficiency $\epsopt - \epsilon_0$ is positive or at least zero in the whole operating area, meaning that the performance is generally improved for every operating point. Especially low-speed operating points benefit from the optimization.

Manipulating the mechanical coefficients shows the greatest impact on the UDC and thus on the performance resulting into a bigger PM and a less robust configuration. Moreover, it has been indicated that a robust optimization is essential to obtain a robust PM configuration when uncertainties play a role. To perform such a robust optimization, the proposed method in this work has proved to be computationally efficient.

\begin{footnotesize}
	\section*{Acknowledgment}
	This work is supported by the German BMBF in the context of the SIMUROM project (05M2013) and the PASIROM project (05M18RDA), by the DFG (SCHO1562/3-1), by the 'Excellence Initiative' of the German Federal and State Governments and the GSCE at TU Darmstadt (GSC 233/2).\par
\end{footnotesize}
%%%%%%%%%%%%%%%%%%%%%%%%%%%%%%%%%%%

\end{document}